\documentclass[12pt,a4paper]{article}
\usepackage{amsmath}
\usepackage[dvips]{graphicx}
\usepackage{times}

\begin{document}
\begin{titlepage}
\title{Reflective scattering effects in  double-pomeron exchange processes}
\author{ S.M. Troshin,  N.E. Tyurin\\[1ex] \small\it Institute
\small\it for High Energy Physics,\\\small\it Protvino, Moscow Region, 142281 Russia}
\date{}
\maketitle
\begin{abstract}
We discuss energy dependence of rapidity gap survival probability in the
double-pomeron exchange processes with account   of
the reflective scattering effects.\\[2ex]
\end{abstract}
\vfill
\end{titlepage}
\section*{Introduction}

Double-pomeron exchange processes are the subject of the
active theoretical studies since the beginning of the nineties,
when the possibility to discover the Higgs production in these
processes was discussed in the seminal papers \cite{dok,land,bj}.
Nowadays these studies are motivated by the current experiments at
Tevatron \cite{tev} and planning for the future experimental
studies at the LHC \cite{ptdr}. Needless to say that  the
processes with large rapidity gaps are important  tool in the
search of new physics. The extensive studies of the magnitude and
energy dependence of gap survival probability have been performed,
results of these studies together with the estimates of the Higgs
production cross--section can be found, e.g. in papers
\cite{khoze,flet,gots,halz,kaid,petr}. Recent results and review
can be found in \cite{rev}.

Perturbative  QCD calculations can be used for calculations of the
cross-sections of such processes where hard scale such as a large
mass is present. But the necessity to take into account soft
rescatterings  in  initial  and final states leads to
uncertainties related to the absence of the theoretically well
established procedure of the multiparticle amplitude unitarization
which is commonly reduced to the absorptive corrections due to the
rescattering processes. Dynamics of such interactions in hard
diffraction processes is accounted then by  introduction of a
factor which is known as a gap survival probability \cite{bj},
i.e. a probability that two colliding particles pass through
without any interaction.

The gap survival probability (SP) was denoted as $\langle |S|^2\rangle$ since it was defined by the
following relation \cite{bj}:
\begin{equation}\label{gap}
\langle |S|^2\rangle=\frac{\int_0^\infty D_H(b)|S(s,b)|^2 d^2b}{\int_0^\infty D_H(b)d^2b},
\end{equation}
where $D_H(b)$ is the probability to observe a specific hard
interaction in collision of the hadrons $h_1$ and $h_2$ and
$|S(s,b)|^2$, where $|S(s,b)|^2=\exp[-\Omega(s,b)]$,  is a
probability that colliding  particles do not have any interaction.
The above definition of SP proceeds from the eikonal scattering
picture \cite{chy} which is usually used for the estimates of
$\langle |S|^2\rangle$, where the eikonal function is denoted by
$\Omega(s,b)$.
 It implies that the soft scattering picture at very high
energy is an absorptive one ($S\geq 0$) and the elastic scattering amplitude in the impact parameter
representation does not exceed  black disk limit which is a half of the unitarity
limit. It is worth to note that the diffractive processes
(with exclusion of elastic scattering)
are  inelastic processes themselves and therefore give contribution to absorption.
Account for the soft processes is model dependent and strongly depends on the
details of the particular model. Estimates for the gap survival probabilities can vary in the wide
range, e.g. from $0.4$ to $22$ \% \cite{miller}. Of course, such wide variation
of theoretical predictions
is not encouraging  for the experimental studies.

In the eikonal picture $S(s,b)$ can easily be treated as the
elastic scattering $S$-matrix, however there are theoretical
problems with the definition given by Eq. (\ref{gap}) and with such
interpretation of the function $S$ if one goes beyond the eikonal
scattering picture . As it was shown in \cite{reflect}, at very
high energies the elastic scattering can go beyond the black disk
limit \cite{phl93} and become a reflective  at small impact
parameters, i.e. elastic $S$ matrix can have negative values. The
generic geometrical
 picture at fixed energy beyond the black disc limit can be described
in this case as a scattering off
the partially reflective and partially absorptive disk
surrounded by the black ring which becomes grey at larger values of the
impact parameter.  The evolution with energy  is characterized
by increasing albedo due to the  interrelated  increase of reflection
  and decrease of absorption at small impact parameters.
Asymptotically, picture of  particle collisions with small impact parameters
looks like collisions of hard balls.
This picture implies
  that the scattering amplitude at
  the LHC energies is beyond the black disk limit at small impact
  parameters (elastic $S$-matrix is negative)
  and it  provides  explanation for the regularities
  observed in cosmic rays studies \cite{reflect}. The straightforward
use of Eq. (\ref{gap}) \cite{rgap} in this case leads to the
non-monotonic energy dependence of SP and its increasing
dependence. It approaches  unity at very high energies which
has no physical meaning. This limit is merely a  manifestation
of the pure reflective elastic scattering ($S\to -1$) at
$s\to\infty$ and $b=0$.
 The reflective elastic scattering does not result from absorption
and therefore the definition given by Eq. (\ref{gap}) cannot be
applied for the $S$-matrices having reflective scattering mode.
The limitation of the definition Eq. (\ref{gap}) to
absorptive scattering is absent if one uses the more general
definition for the SP (we denote this quantity as $R_s^2$ to
differentiate it from the Eq. (\ref{gap})) given in
\cite{miller,bart,strikman}
\begin{equation}\label{gapn}
R_s^2(s)=\frac{\sigma_{diff}(\mbox{phys})}{\sigma_{diff}(\mbox{hard})},
\end{equation}
where $\sigma_{diff}(\mbox{hard})$ can, in principle, be calculated with perturbative QCD, while
$\sigma_{diff}(\mbox{phys})$ is the real cross-section, which can be confronted with the cross-section
measured experimentally. The definitions Eqs. (\ref{gap}) and (\ref{gapn}) are equivalent if $S$
is nonnegative in the whole range of variation of the variables $s$ and $b$. However, they are not
equivalent if $S$ can become negative at some values of the energy and impact parameter.

We  use therefore the definition Eq. (\ref{gapn}) and calculate the energy
dependence of $R_s^2(s)$ in the unitarization scheme which includes reflective scattering in the
double-pomeron exchange processes (DPE).

Now a few comments are to be added. First of all, of course, the quantity $R_s^2$ is model and
process dependent. Another important point was mentioned in \cite{halz}, this quantity is probability
of surviving rapidity gaps, it is not the probability of production and surviving rapidity gaps.
It is useful to have it in mind under discussion of the available experimental data on hard hadronic diffraction
at Tevatron.

Experimental studies at Tevatron  energies $\sqrt{s}=630$ GeV and $\sqrt{s}=1800$ GeV
show that in the hard double-diffractive (DD) interactions
 the ratio of the  of the corresponding
 event numbers decreases with energy \cite{tev}. The studies of  inclusive DD at CDF \cite{tev}
 leads to conclusions on the universality of SP in soft and hard diffractive processes and on the absence
 of suppression of the additional gaps compare to the diffractive processes with single rapidity gap, i.e.
 SP does not depend on number of gaps.

\section*{SP in double-pomeron exchange processes; effects of the reflective scattering}

 As it was already mentioned, almost all estimates of  SP are performed on the basis
  of the eikonal amplitude
  unitarization scheme. To strengthen   the obtained predictions it is desirable to have predictions
  based on the other forms of the amplitude unitarization. It is important to move this way
   in the situation when there is no
  unique  scheme for the unitarity account.

 Alternative form of unitarization uses the rational representation
  for the scattering amplitude.
The rational form of unitarization in quantum field theory
 is based on the relativistic generalization \cite{umat}
 of the Heitler equation \cite{heit}.
In this $U$--matrix approach the elastic scattering matrix in the impact
parameter representation has the form:
\begin{equation}
S(s,b)=\frac{1+iU(s,b)}{1-iU(s,b)}, \label{um}
\end{equation}
where $S(s,b)=1+2if(s,b)$ and $U(s,b)$ is the generalized reaction matrix, which is
considered to be an input dynamical quantity similar to the
eikonal function. Unitarity equation rewritten at high energies
for the elastic amplitude $f(s,b)$ has the form
\begin{equation}
\mbox{Im} f(s,b)=|f(s,b)|^2+\eta(s,b) \label{unt}
\end{equation}
where the inelastic overlap function
\[
\eta(s,b)\equiv\frac{1}{4\pi}\frac{d\sigma_{inel}}{db^2}
\]
 is the sum of
all inelastic channel contributions.
Inelastic overlap function
is related to $U(s,b)$ according to Eqs. (\ref{um}) and (\ref{unt}) as follows
\begin{equation}
\eta(s,b)=\frac{\mbox{Im} U(s,b)}{|1-iU(s,b)|^{2}}\label{uf},
\end{equation}
i.e.
\begin{equation}\label{sinel}
\sigma_{inel}(s)=8\pi\int_0^\infty bdb \frac{\mbox{Im} U(s,b)}{|1-iU(s,b)|^{2}}.
\end{equation}
It should be noted that
\begin{equation}\label{imu}
\mbox{Im} U(s,b)=\sum_{n\geq 3} \bar U_n(s,b),
\end{equation}
where $\bar U_n(s,b)$ is a Fourier--Bessel transform of the function
\begin{eqnarray}\label{un}
\bar U_n(s,t) & = & \frac{1}{n!}\int \prod_{i=1}^n\frac{d^3q_i}{q_{i0}}\delta^{(4)}(\sum_{i=1}^n q_i-p_a-p_b)
U_n^*(q_1,....,q_n;p_{a}',p_{b}')\cdot\\
& &U_n(q_1,....,q_n;p_a,p_b)\nonumber.
\end{eqnarray}
Here the functions $U_n(q_1,....,q_n;p_a,p_b)$ and $U_n(q_1,....,q_n;p_{a'},p_{b'})$ correspond to the
ununitarized (input or ``Born'') amplitudes of the process
\[
a+b\to 1+....+n,
\]
and the process with the same final state and the initial state with different momenta $p_a'$ and $p_b'$
\[
a'+b'\to 1+....+n,
\]
respectively. They are the analogs of the elastic $U$-matrix
for the processes $2\to n$. The sum in the right hand side of the
Eq. (\ref{un}) runs over all  inelastic final  states $|n\rangle$
 which include  diffractive  as
well as non-diffractive ones. Graphically, these relations are illustrated in Fig. 1.
\begin{figure}[hbt]
\begin{center}
\includegraphics[scale=0.7]{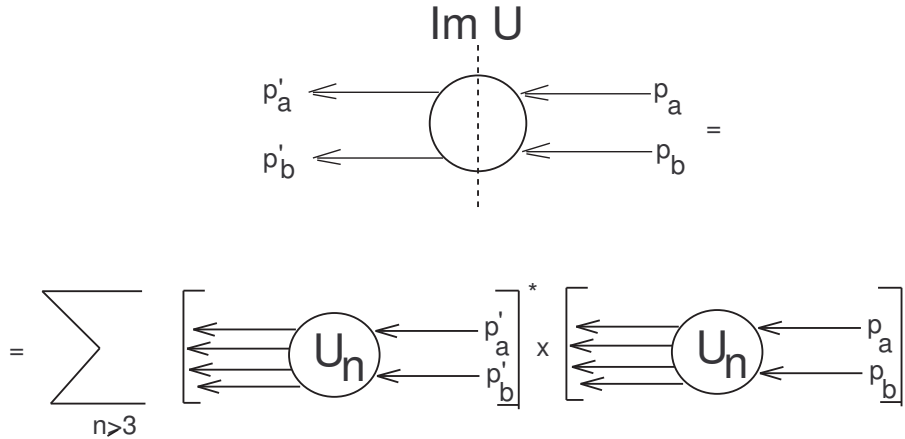}
\caption{{\it Unitarity  for the function} $\mbox{Im} U$.}
\end{center}
\end{figure}

Then the inclusive cross-section of the process $ab\to cX$ has the following form
\cite{tmf}, which is similar to the expression for the total inelastic
cross--section (\ref{sinel}):
\begin{equation}
E\frac{d\sigma}{d^3q}= 8\pi\int_0^\infty
bdb\frac{I(s,b,q)}{|1-iU(s,b)|^2}\label{unp},
\end{equation}
where $I(s,b,q)$ is the Fourier-Bessel transform of the functions which are defined
similar to Eq. (\ref{un}) but with the fixed momentum $q$ and energy $E$ of the particle $c$ in
 the final state (Fig. 2).
\begin{figure}[hbt]
\begin{center}
\includegraphics[scale=0.7]{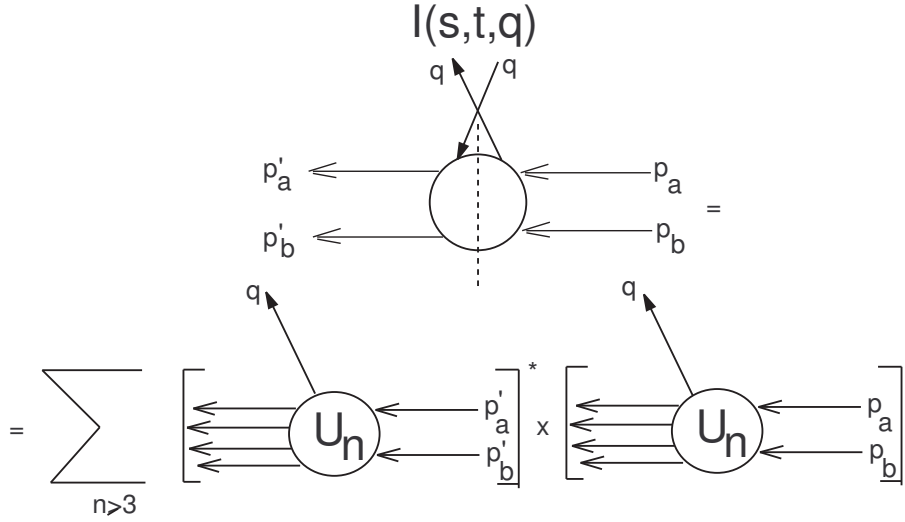}
\caption{{\it Unitarity  for the function} $I(s,t,q)$.}
\end{center}
\end{figure}
It should be noted that the impact parameter $b$ is the variable conjugated to the variable $\sqrt{-t}$, where
$t=(p_a-p_a')^2$ and that the following sum rule is valid for the $I(s,b,q)$
\begin{equation}\label{sum}
\int \frac{d^3q}{E}I(s,b,q)=\bar n(s,b)\mbox{Im} U(s,b).
\end{equation}
The impact parameter $b$ is related to the impact parameters of the
secondary particles by relation \cite{sakai}
\begin{equation}\label{v}
\mathbf{b}=\sum_{i=1}^n x_i\mathbf{b}_i,
\end{equation}
where $x_i$ stands for the Feynman variable $x$ of the $i$-th particle.
We  consider the double-pomeron exchange processes of the
following type
\begin{equation}\label{dpe}
pp\to p+X+p,
\end{equation}
where $+$ denotes gap in rapidity and $X$ can be a particle or a
system of particles. To  obtain the cross-section  of the DPE
process (\ref{dpe}) we should single out in the sum over all
inelastic final states
 $|n\rangle $ the specific final states
 $|n\rangle_{dpe}$, which are the  final  states
corresponding  to  the specific kinematics of the corresponding DPE process.
Then denoting via $\zeta$ a set of kinematical variables which characterizes
the kinematics of the final state in reaction (\ref{dpe}), a similar to Eq. (\ref{unp})
expression can be written for the cross-section $d\sigma_{dpe}/d\zeta$
\begin{equation}\label{crsdpe}
\frac{d\sigma_{dpe}}{d\zeta}= 8\pi\int_0^\infty
bdb\frac{I_{dpe}(s,b,\zeta)}{|1-iU(s,b)|^2}.
\end{equation}
The formula (\ref{crsdpe}) is  consistent with the experimental
result of CDF on the independence of SP on the gap number.
The function $U(s,b)$ is related to the elastic scattering amplitude and can,
in principle, be extracted from the elastic scattering experimental data, however,
in practice, the various models can be used to construct the function $U(s,b)$ and get
its form at the energies where the data are absent at the moment. The models also
needed to construct $I_{dpe}(s,b,\zeta)$. The ratio
\begin{equation}
\int d \zeta I_{dpe}(s,b,\zeta )/
\int \frac{d^3q}{E} I(s,b,q)=c(s,b) \label{produc}
\end{equation}
 should be considered as
a  gap production probability. From the Eqs. (\ref{sum}) and (\ref{produc})
 it follows that
\begin{equation}\label{csb}
\int d \zeta I_{dpe}(s,b,\zeta )=c(s,b)\bar n(s,b)\mbox{Im} U(s,b).
\end{equation}

To get quantitative results we  use earlier constructed model for the $U$--matrix
 based on the ideas of chiral quark models \cite{chpr}.
The picture of a hadron consisting of constituent quarks in the central part
of hadron embedded
 into quark condensate implies that overlapping and interaction of
peripheral clouds   occur at the first stage of hadron interaction.
The impact parameter dependence of this interaction is determined by the function $D_C(b)$,
which is a convolution of
the two condensate distributions $D^{a}_c({b})$ and
$D^{b}_c({b})$ inside the hadrons $a$ and $b$.
We  consider for simplicity the case of a pure imaginary amplitude,
i.e. $U\to iu$.
The
function $u(s,b)$  is represented in the model as a product of the
averaged quark amplitudes $\langle f_Q \rangle$,
\begin{equation} u(s,b) =
\prod^{N}_{i=1} \langle f_{Q_i}(s,b)\rangle \end{equation} in
accordance  with the assumed quasi-independent  nature  of the valence
quark scattering, $N$ is the total number of valence quarks in the
colliding hadrons.  The essential point here
is the rise with energy of the number of the scatterers  like
$\sqrt{s}$. The $b$--dependence of the function
$\langle f_Q \rangle$  has a simple form
$\langle f_Q(b)\rangle\propto\exp(-{m_Qb}/{\xi} )$.
The generalized
reaction matrix  gets
the following  form
\begin{equation} u(s,b) = g\left (1+\alpha
\frac{\sqrt{s}}{m_Q}\right)^N \exp(-\frac{Mb}{\xi} ), \label{x}
\end{equation} where $M =\sum^N_{Q=1}m_Q$.
Here $m_Q$ is the mass of constituent quark, which is taken to be
$0.35$ $GeV$\footnote{Other parameters have the following values:
 $g=0.24$, $\xi=2.5$, $\alpha=0.56\cdot 10^{-4}$.}.
The model provides  linear
dependence on $\sqrt{s}$ for the total cross--sections,
 i.e.
 $\sigma_{tot}=a+c\sqrt{s}$
    in the limited energy
range $\sqrt{s}\leq 0.5$ TeV in agreement with experimental data  \cite{pras,univ}
 and asymptotically,
 $\sigma_{tot}\propto \ln^2 s$, while $\sigma_{inel}\propto \ln s$.

Inelastic diffractive processes in the central region (DPE) occur in the peripheral
region of the impact parameter space. Constituent quarks, which are responsible
for elastic scattering, are supposed to be located  in the central part. This fact and
relations (\ref{produc}) and (\ref{csb}) allow one to assume factorization at the
$U$-matrix level and write down the expression for $I_{dpe}(s,b,\zeta)$ in the form
\begin{equation}\label{idpe}
I_{dpe}(s,b,\zeta)=\sigma_{dpe}(s,b,\zeta)\mbox{Im} U(s,b),
\end{equation}
where $\sigma_{dpe}(s,b,\zeta)$  describes hard or soft  "Born"
cross--section of DPE processes.
It should be  stressed that the factorization (\ref{idpe}) is assumed  to be valid
for both soft and hard  DPE processes at the
$U$-matrix level. We do not expect similar factorization, e.g. for DD
processes. This factorization is different also  from
the factorization discussed in \cite{strikman} at the $S$-matrix level which is considered to be
 valid for all diffractive processes with hard scale.

We consider  hard DPE processes because they are the most
interesting ones, i.e.
\[
\sigma_{dpe}(s,b,\zeta)\to \sigma_{H}(s,b,\zeta).
\]
Thus, under construction of $\sigma_{H}(s,b,\zeta)$ we will suppose that the
interaction occur at small distances. We can write down then
 the probability of hard interactions in the model as a convolution
\begin{equation}
\sigma_{H}(s,b,\zeta)=\int D_c^{h_1}({\bf b}_1)\sigma^0_H(s, {\bf b}+{\bf b}_1-{\bf b}_2, \zeta)D_c^{h_2}({\bf b}_2)
d{\bf b}_1 d {\bf b}_2,
\end{equation}
where   $\sigma^0_H(s, {\bf b}+{\bf b}_1-{\bf b}_2, \zeta)$ is the cross--section of the
hard condensate (parton) interactions. If the interaction is hard, than it is natural to choose
\[
\sigma^0_H(s, {\bf b}+{\bf b}_1-{\bf b}_2, \zeta)\simeq \sigma^0_H(s, \zeta)
\delta^2({\bf b}+{\bf b}_1-{\bf b}_2).
\]
It means that hard interaction cross-section can be represented as a convolution integral
\begin{equation}\label{cyang}
\sigma_{H}(s,b,\zeta)\simeq\sigma^0_H(s, \zeta)\int D_c^{h_1}({\bf b}-{\bf b}_1)D_c^{h_2}({\bf b}_1)
d{\bf b}_1,
\end{equation}
which is similar to the  model originally proposed by Chou and Yang \cite{chy}.
This model can be considered, in fact, as a model based on hard parton interactions.
It is useful to note here that the hardness of the inelastic interaction does not imply the smallness
 of the impact parameter ${\bf b}$ (cf. Eq. (\ref{v})),
 the typical values of the impact parameter $b$ are determined
 by the  condensate distributions $D_c^{h_1}$ and $D_c^{h_2}$ which are
 responsible for the central diffractive production
 and have mostly peripheral impact parameter dependence. The two-scale
 picture of hadron scattering proposed in \cite{strikman,islam}, where the core  of hadron is responsible
 for hard interactions and peripheral part -- for soft interaction, should definitely
  take place in elastic diffraction, while the hard interaction in the inelastic diffraction
 can take place in the peripheral region as well.
 Of course, the energy dependencies of soft and hard hadron interaction radii are different,
 the hard interaction
 radius of hadrons is energy independent, while the soft hadron interaction radius
 rises with energy as $\ln s$, due to this difference in the energy dependencies
 the limit $R^2_s(s)\to 0$ at $s\to \infty$ takes place.

 After these qualitative remarks we would like to provide a quantitative estimates for the energy
 dependence of $R^2_s(s)$ for DPE processes. We adopt a most
 simple choice and assume that the condensate distributions
 $D_c^{h_i}$ over impact parameter $b$ is controlled by the pion mass, then the convolution
 \begin{equation}\label{conv}
D_c^{h_1}\otimes D_c^{h_2}\sim \exp(-m_\pi b).
\end{equation}
In Eq. (\ref{cyang}) we do not write down explicitly dependence
on the parton momentum fraction $x$; one can suppose that the $x$
and $b$ dependencies are factorized (at the $U$--matrix level) and
give contribution to the cross-section $\sigma^0_H(s, \zeta)$ which
will be cancelled out in the dependence of $R^2_s(s)$. Now we
calculate this dependence with the  models for the function $U(s,b)$
and for the impact parameter dependence of the function
$\sigma_{H}(s,b,\zeta)$.
According to Eq. (\ref{conv}) we should calculate the ratio of the two integrals
$\sigma_{H}(s,b,\zeta)\eta(s,b)$ and $\sigma_{H}(s,b,\zeta)$ over the variables $b$ and $\zeta$.
The function $R^2_s(s)$ in the  range from the Tevatron energies till the LHC energies and beyond
this region is depicted on Fig. 3.
\begin{figure}[hbt]
\begin{center}
\includegraphics*[scale=0.4]{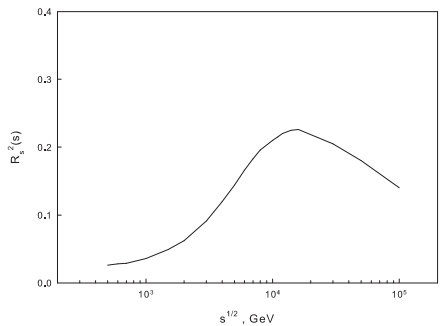}
\caption{{\it Energy dependence of survival probability $R_s^2(s)$.}}
\end{center}
\end{figure}

Of course, numerical predictions for the gap survival probability
 obtained here are model dependent, but the qualitative picture of the
 energy dependence of $R^2_s(s)$ reflects transition to the new
 scattering mode at the LHC energies.
  It is interesting that situation is more favorable at the LHC energies
 since the obtained numerical values of $R^2_s(s)$ at these
  energies are close to maximal values around 0.2 and this should lead to higher cross--sections
e.g. for Higgs production in DPE processes compared to the
   values obtained with eikonal estimations of the gap survival probability.
Thus, appearance of the reflective scattering  would result in better
 detecting of Higgs boson in DPE processes.
\section*{Conclusion}
We would like to note  that Eq. (\ref{gapn}) is a more general definition
of the gap survival probability than  Eq. (\ref{gap}). Both are equivalent in the
case of absorptive scattering, but they are different  when the reflective scattering is present.
The relevant definition in the latter case is Eq. (\ref{gapn}).

The use of Eq. (\ref{gapn})
leads to the value of $R^2_s(s)\sim 0.2$ for the hard DPE processes at the LHC energies instead of
unity which has been obtained in \cite{rgap} on the base of the definition Eq. (\ref{gap}).
The energy dependence of $R^2_s(s)$ is determined by the interplay of soft and hard interaction
radii. At $s\to\infty$ the ratio $R^2_s(s)\to 0$ due to dominance of the soft hadron interaction radius
over the hard interaction radius of hadrons.

The LHC energies appear to lie in the favorable energy region, where soft and hard hadron interaction radii
are close and this allows one to arrive to rather optimistic conclusions on the Higgs finding in the
DPE processes there.
\section*{Acknowledgement}
We are grateful to M. Arneodo, M. Grothe and V. Petrov for the useful discussions on the subject
of the letter.

\end{document}